\documentclass[referee]{raa} 
\usepackage{graphicx,times}
\usepackage{natbib}
\usepackage{amssymb,amsmath}
\bibpunct{(}{)}{;}{a}{}{,}

\usepackage{graphicx}
\usepackage{dcolumn}
\usepackage{bm}
\usepackage{hyperref}
\hypersetup{pdftitle = The title of my PDF, pdfauthor = My name, pdfsubject= The subject, pdfkeywords = keyword1 keyword2 keyword3} 
\hypersetup{colorlinks = true, linkcolor = green, anchorcolor = red, citecolor = blue, filecolor = red, pagecolor = red, urlcolor = red}

\newcommand{\Code}{PHoTo{\it N}s }

\begin{document}

\title{ \Code--A Parallel Heterogeneous \& Threads oriented code for cosmological {\it N}-body Simulation\textbf{}}

\volnopage{ {\bf 2018} Vol.\ {\bf X} No. {\bf XX}, 000--000}
\setcounter{page}{1}

\author{Qiao Wang\inst{1}, Zongyan Cao\inst{2,1}, Liang Gao\inst{1,3}, Xuebin Chi\inst{2}, Chen Meng\inst{2,4}, Jie Wang\inst{1}, Long Wang\inst{2,4}}

\institute{
Key Laboratory of Computational Astrophysics, National Astronomical Observatories, Chinese Academy of Sciences, Beijing, 100012 {\it qwang@nao.cas.cn}\\
\and
Supercomputing Center, Computer Network Information Center, Chinese Academy of Sciences, Beijing\\
\and
Institute for Computational Cosmology, Department of Physics, Durham University, Science Laboratories, South Road, Durham Dh1 3LE, England \\
\and
System Department, Baidu Group Ltd., Beijing \\
\vs \no
{\small Received 0000 xxxx 00; accepted 0000 xxxx 00}
}

\abstract{
We introduce a new code for cosmological simulations, \Code, which has features on performing massive 
cosmological simulations on heterogeneous high performance Computer (HPC) and threads oriented programming. \Code adopts a hybrid scheme to compute gravity force, with the conventional PM to compute the long-range force, the Tree algorithm to compute the short range force, and the direct summation PP to compute the gravity from very close particles. A self-similar space filling Peano-Hilbert curve is used to decompose computing domain. Threads programming is highly used to more flexibly manage the domain communication, PM calculation and synchronization, as well as Dual Tree Traversal on the CPU+MIC platform. The scalability of the \Code performs well and the efficiency of PP kernel achieves 68.6\% of peak performance on MIC and 74.4\% on CPU platforms. We also test the accuracy of the code against the much used Gadget-2 in the community and found excellent agreement.
}

\authorrunning{Q. Wang et al. }  
\titlerunning{PHoTo{\it N}s}
\maketitle

\section{Introduction}
\label{sec:introduction}

During the last decades, cosmological N-body simulation becomes an essential tool to understand the large scale structure of the Universe, due to the nonlinear physical nature of the formation and evolution of cosmic structure. Demanded by future large scale galaxy surveys, the new generation of cosmological simulation requires a huge number of particles and a huge simulation volume in order to obtain good statistics and simultaneously achieve high resolution to resolve faint galaxies to be observed in galaxy surveys. For example, a recent simulation of ~\citet{2017ComAC...4....2P} evolving unprecedented 2 trillion particles in 3 Gpc$^3$ volume to make galaxy mock for the upcoming Euclid survey. To perform such simulation, an exquisite simulation code is a must. Indeed many cosmological simulation codes have been continuously developed in recent years ~\citep{1985ApJS...57..241E,1991ComPh...5..164B,2002ApJ...574..538J,2002A&A...385..337T,2003PASJ...55.1163M,2004PASJ...56..521M,2004NewA....9..137W,2005MNRAS.364.1105S,2009MNRAS.398L..21S,2009MNRAS.396.2211B,2009PASJ...61.1319I,2011ApJ...740..102K,2012MNRAS.423.3018P,2012arXiv1211.4406I,2013arXiv1310.4502W,2015MNRAS.450.4070W,2017RAA....17...85E,2017NatAs...1E.143Y}.

At the mean time, the high performance computing architecture is undergoing significant changes, many new HPC systems have mixed CPU+GPU architecture rather than the traditional pure CPU one~\citep{ 1997ApJ...480..432M, 2003PASJ...55.1163M}. Many existing cosmological simulation codes were developed based on the pure CPU systems and thus are not able to take advantage of the power of the popular heterogeneous HPC machines~\citep{2011MNRAS.413..101G}. In this paper, we introduce a novel cosmological simulation code \Code. It has two characteristics, the first feature is that the code is programmed basing on threads, the other is highly optimized on heterogeneous architecture, such as Intel Xeon Phi (MIC). 

The organization of the paper is as follows. We briefly introduce the background of physical model in section 2.  The force calculation algorithm is present in Section 3. Section 4 discuss our parallelization and implementation strategy. In section 5, we describe the numerical accuracy and performance of the \Code.

\section{Equations of Motion}
\label{sec:equation}
In cosmological N-body simulation, the underlying matter of the Universe is usually sampled with collision-less dark matter particles which are only governed by gravity. Thus the Lagrangian of a particle reads 
\begin{equation}
\mathcal{L} = \frac{{\mathbf p}^2}{2a^2} - \frac{\phi}{a},
\label{eq:lagrangian}
\end{equation}
where the momentum ${\mathbf p} = a{\mathbf v}$, $\phi$ refers to comoving potential ~\citep{1980lssu.book.....P}, which is determined by Poisson's equation
\begin{equation}
\nabla^2 \phi = 4\pi G a^3 \left[\rho_m\left({\mathbf r},t\right) - {\bar \rho_m}\left(t\right)\right],
\label{eq:poisson}
\end{equation}
 The density field $\rho$ can be constructed from discrete mass points
\begin{equation}
\label{eq:dens}
\rho \left({\mathbf r}\right) = \sum_{\forall i} \frac{m_i} {a^{3}} \delta_{\bf{D}}\left({\mathbf r}-{\mathbf r}_i\right). 
\end{equation}
Here the Dirac delta function $\delta_D$ can be modified according to different mass distributions.

According to the Lagrangian, the motion equations of an individual particle are given by
\begin{eqnarray}
\label{eq:eom}
\dot{\mathbf r} &=&  \frac{{\mathbf p}}{a^2}, \notag \\ 
\dot{\mathbf p} &=& -\frac{\nabla \phi}{a},
\end{eqnarray}
It is apparent that the gravity between two particles satisfies the inverse-square law due to the form of the Poisson equation. The acceleration of a particles is solved by the gradient of potential and pairwise Newtonian gravity in the simulation.

\section{Force calculation Algorithm \& implementation}
The most challenging problem for cosmology N-body simulation is how to compute the gravitational force efficiently and accurately. In \Code we adopt a hybrid scheme to do the task, namely using Particle-Mesh algorithm to compute the long range force, a novel combination of Tree algorithm and direct summation Particle-Particle(PP) to compute the short range force. We  described our force calculation scheme in details below. 

\subsection{PM-Tree-PP method} 

The Particle-Mesh (PM) method ~\citep{1988csup.book.....H} is a common and efficient algorithm to solve the Poisson equation (Eq.~\ref{eq:poisson}). In Fourier space, the Poisson equation can be expressed as an algebraic equation, its solution can be simply obtained by convolution of Green function under the periodic boundary condition. The Green function of the Poisson equation has a simple form $-/k^2$, its 3-point difference in $k$  space reads
\begin{equation}
g_k({l,m,n}) = \frac{\pi G \Delta_g^2}{\sin^2 \left( \frac{\pi }{N} l\right)+{\sin}^2 \left(\frac{\pi }{N} m\right)+{\sin}^2 \left(\frac{\pi}{N} n\right)}
\end{equation}
where $\Delta_g$ denotes the width of a mesh, $N$ is the mesh number on one side, $l, m, n$ denotes the discrete wave number. The convolution of density field and Green function is a multiple in Fourier space and can be implemented with Fast Fourier Transformation(FFT).  As FFT works on regular mesh, one needs to assign the density filed sampled by a number of discrete particles into a regular mesh. Then the Poisson equation is solved in Fourier space, and an inverse FFT transformation is applied to obtain gravitational potential on meshes. Finally the acceleration of each particle can be linearly interpolated from the potential meshes. 

The force accuracy of the PM method is nearly exact at large scales but drops dramatically within a few mesh sizes \citep{2002JApA...23..185B}. In past years many others gravity solver algorithms were invented to improve force accuracy in the small scales. Among these, a popular gravity solver is {\it Tree method} invented by ~\citet{1986Natur.324..446B}, which provides a robust efficiency even for highly clustering system. In their original approach, the simulation volume is recursively divide into 8 smaller subcubes until each subcube only contains one particle. Each subcube contains information of the center of mass and the multiple moment for the mass distribution enclosed in the subcube etc. Considering every particle as a leaf, one can organize the adjoined particles in hierarchical branches (tree nodes) of a tree data structure. So that the gravity from distant tree nodes can be computed as individual mass points, otherwise the closer nodes are opened. The force calculation of a particle is complete when the tree-walk recursively go though all branches. The time complexity of such a Tree method is ${\mathcal O}(N{\log}N)$.

\citet{2002JApA...23..185B} suggests a hybrid approach to combine the advantage of the Tree and PM methods in order to achieve adequate force accuracy at both large and small scales. The idea is to elaborately compute the long range gravity with the PM method and calculate the short range force with the Tree algorithm. The combination of two parts achieves an accurate gravity at all scales. To this end, for each particle, when doing the PM calculation, an additional low-pass Gaussian filter, $1-\exp(-s^2/2R_s^2)$, is convoluted to take out the short range force of the PM calculation. Here $s$ is the spatial separation of two particles,  $R_s$ is a characteristic parameter to control the splitting scale at which the long and short range force calculation take effect. Correspondingly, the expression of the short range force also need to be modified by the following factor

\begin{equation}
\label{equ:split}
\mathbf{f}^{sh}_i({s}) =  {\rm erfc}\left( \frac{s}{2R_s} \right) + \frac{s}{R_s\sqrt{\pi}} {\rm exp}\left( {-\frac{s^2}{4R_s^2}} \right) ,
\end{equation}
where $s$ is the spatial separation from the particles $i$. $R_s$ is the splitting radius. From the expression one can readily see that the short range force drops rapidly as the separation increases,  beyond a certain scale $R_{cut}$, contribution from the short force is negligible. Here exp and erfc functions in  Eq.~\ref{equ:split} both are computationally expensive,  and so Eq.~\ref{equ:split} is usually estimated based on the interpolation from a pre-built -$float$ table.

Following \cite{2002JApA...23..185B} and ~\citet{2005MNRAS.364.1105S}, \Code also carries out the most force calculation with PM-Tree scheme \citep{1995ApJS...98..355X}. For the PM part, \Code employs a conventional Cloud-in-Cell (CIC) to assign particles into regular grids. For the Tree calculation part, we adopt a much used  oct-tree~\citep{1986Natur.324..446B} for the tree construction. In contrary to some existing Tree codes, we follow  ~\citet{2005MNRAS.364.1105S} to adopt monopole moments for tree nodes. As discussed in ~\citet{2005MNRAS.364.1105S}, there are some attractive advantages in using the monopole moments scheme, for example, less memory consumption which also improves the efficiency of tree operations. But the monopole scheme usually requires a more strict opening  criterion in order to achieve the same level of force error when comparing to multiple moment. In addition to the geometric effect, the accuracy of tree method is also affected by dynamical state, we follow the criterion of Gadget-2,
\begin{equation}
\frac{GM}{s^2}\left( \frac{l}{s}\right)^2 \le \alpha |{\bf a }|,
\end{equation}
where $s$ is the separation between the target particles, tree node with the width of $l$, $\alpha$ is a control parameter, and ${\mathbf a}$ is acceleration of particle. this opening criterion results in higher force precision for particles with larger acceleration and is more effective than pure geometric opening criterion, please refer to ~\citet{2005MNRAS.364.1105S} for details.

When considering the scale at which the PM and Tree calculation are split, \citet{2002JApA...23..185B} found that the gravity error is less than 1\% if $R_{cut}$ is 3.5 times larger than $R_s$. A more strict cut scale is used in \Code, which follows Gadget-2 \citep{2005MNRAS.364.1105S}, $R_{cut} = 4.5 R_s$ and $R_s = 1.2 \Delta_g$. ~\citet{2005MNRAS.364.1105S} shows that, with these parameters, the force error of more than 99.99\% particles is smaller than 0.005\% for a typical value of $\alpha=0.001$. It implies the $R_{cut}$ should be larger than $5.4 \Delta_g$. Thus our choice of $R_{cut}=6 \Delta_g$ is sufficient.

\begin{figure}[htbp]
\centering
\includegraphics[width=0.6\textwidth]{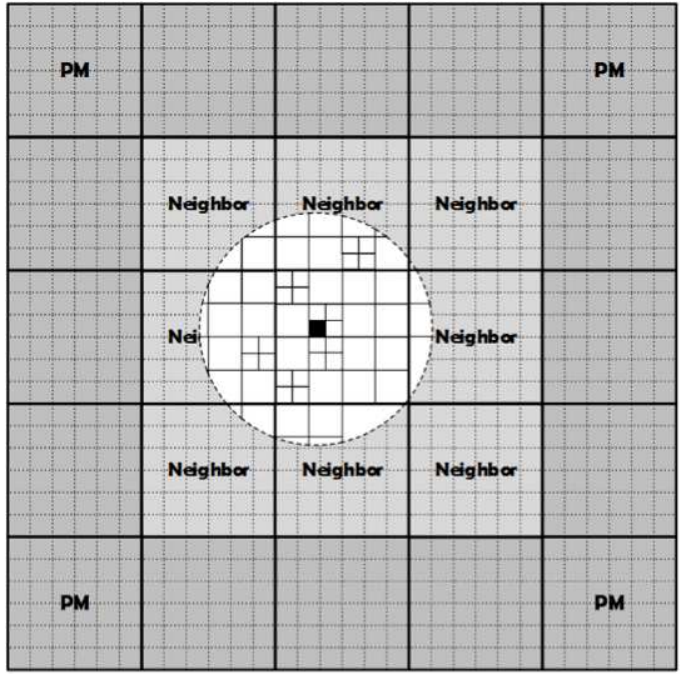}
\caption{Diagram for the long and short range force decomposition, the coarse grain corresponds to the ground tree node and fine grid is for the PM calculation. Gravitational force in any target area (the filled square) is a summation of the long range PM and short range Tree force. The radius of dash circle is exactly 6 times the PM grid, thus the adjoining ground tree node (coarse grain) contains all information needed for the short range force calculation.}
\label{fig:split}
\end{figure}

We demonstrate the idea of the force splitting in Fig.~\ref{fig:split}. The coarse grains (heavy solid line) are referred to as the $ground$ tree-node and the number of one side (N$_{side}$) should be 2 to the power of n. The fine grains (dotted line) represents the PM mesh. The size of the coarse grains is set to be exactly 6 times the fine one ($> 5.4 \Delta_g$), which has advantage that the tree calculation part for the target point (a filled square in the diagram) only needs to consider its adjacent coarse grains and itself. Our domain decomposition strategy is also based on this splitting scheme, we will discuss this further in later sections.

In the standard PM-Tree method, the short range force is computed all the way with the tree method. \Code makes some modifications, instead we use the PP method to replace the tree calculation when a tree branch left a given number of particles, $N_{pp}$. This implementation has two advantages. Firstly by doing this the depth of the tree is reduced by a factor of about $15-20\%$ hence reduces the memory consumption and the levels of tree-walk. Secondly, compared to the Tree method, it is more efficient to compute force of a group particles with the PP if the number is small. In practice, we found $N_{pp}=100$ is a good number to keep the force calculation efficiency and at the same time reduce memory consumption.\Code also made several improvements on the Tree-building part as described below. 

\subsection{Tree building}
\label{sec:treebuild}

\Code initially builds a conventional oct-tree as other Tree codes does, namely a series of pointers are defined in the tree nodes structure to record the relationship of parent and off-spring nodes. During the procedure, tree is built recursively by inserting particles one by one, as a result tree nodes are not contiguously ordered in memory, which is not friendly for tree-walk. Some modifications are made to improve the tree-walk efficiency by rearranging the tree nodes later in the order of tree-walk, however even by doing this the data storage for particles is still in-contiguous. Such a scheme is neither friendly for memory access nor hardware optimization, and it is even more serious on some heterogeneous architecture, for instance Intel MIC due to communication and memory allocation strategy on MICs.

\begin{figure}
\centering
\includegraphics[width=0.9\textwidth]{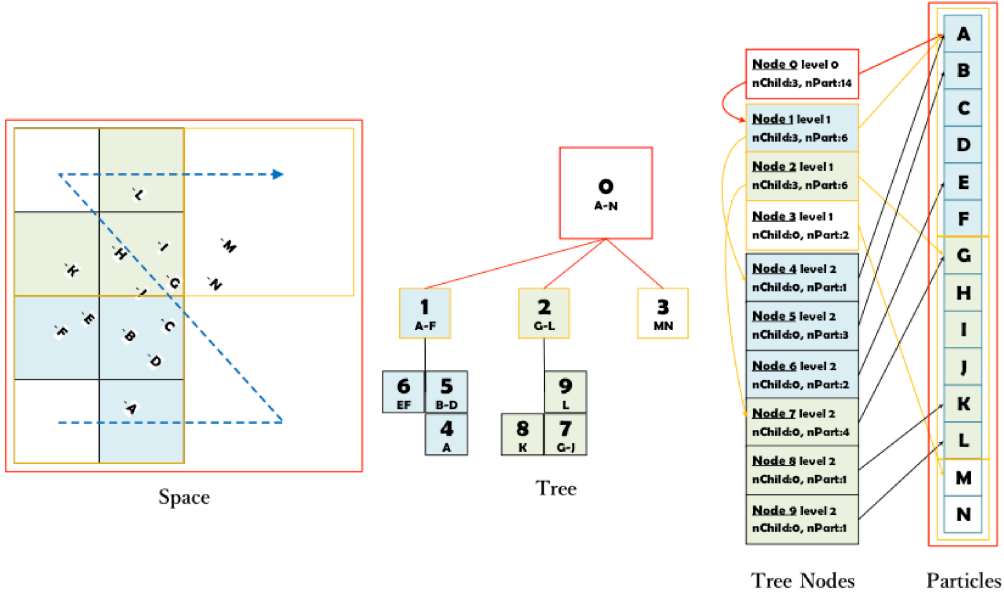}
\caption{A diagram for the tree structure. The left panel displays a group of particles (from A to N) residing in a cube on the configuration space, The middle panel show the tree structure, and the right panel displays these 14 particles and 10 tree nodes are stored in the contiguous memory after operations, see text for details.}
\label{fig:treemic}
\end{figure}

Desirably, all particles of the same tree node should be in a contiguous sequence in memory, \Code adopt two steps to achieve this . Firstly all $ground$ grids are marked with the index of $( i \times $N$_{side} + j ) \times$N$_{side} + k$. Then we label each  particle with the index of the grid in which it resides and apply for a bucket sorting to collect the particles with the same index into the same grid.  Now all particles in the same grid should be adjacent in the $ground$ level. In the second step we need to refine the $ground$ tree nodes to make them continuous in memory. We adopt Morton key to re-arrange all particles of each grid. Morton key is a spatial filling sequence, its order of the key is exactly the same as the structure of the oct-tree. After doing this, all particles are assigned into the proper tree node at each level, and the number of the particle in a node and the position (offset) of the first particle in the global array are recorded in the tree node data structure.  After all these steps, not only the children of a branch node are contiguously stored in the memory, but also a particle in any node can also be easily identified in just one memory block without much time latency caused by the cache miss. This scheme greatly improves the memory access efficiency in consequent tree-walk procedure, especially on MIC. The memory arrangement of tree and particle is shown in 2D in Fig.~\ref{fig:treemic}.

\subsection{Dual Tree Traversal}
\Code Allows an option to run on Intel MIC architecture by employing a Dual Tree Traversal (DTT) scheme to calculate gravity force. In the conventional Tree method, the gravity force of a target particle is the interaction between the target particle and tree nodes and so needs to walk through the entire tree, while the DTT computes gravity based on interactions between tree nodes as described in details below. 

Given the side width of the two nodes, $A$ and $B$ respectively, we define the two opening criteria $\theta_1 = (B + A)/R$, and $\theta_2 = B/(R - A)$. If the opening angle is larger than the criterion, such as $\theta=0.3$, one node should be opened and its off-springs are checked respected to the other node recursively, until all node pairs are accepted or end up as a leaf \citep{lange2014}. In our framework, a leaf is a package of a given number of particles. Once the criterion meet, all particles of the target node gain the gravity from the mass center of the other node (in the case of 1 to N). If traversal occurs between two fat leaves (in the case of N to M), a subroutine of PP kernel is called, this case usually occurs in dense regions. Since particles are contiguous in memory after the tree building as we described in the last section, the force accumulation can be readily optimized. 

Regards to the opening criterion, previous study \citep{lange2014} suggest to open the relatively bigger node\citep{lange2014}. In our practice, however, we found this is a good choice only on pure CPU platform. On MIC architecture, it performs better to preserve the target node and open the other one, no matter which one is larger. 

\subsection{adaptive KDK stepping}
\label{sec:step}
The N-body problem is a Hamiltonian system whose long time behavior can often be changed by non-Hamiltonian perturbations introduced by ordinary numerical integration methods, thus a  symplectic integration method is desired. Here we adopt the cosmic symplectic scheme proposed by \cite{1997astro.ph.10043Q}. Based on the Lagrangian of Eq.~\ref{eq:lagrangian}, two symplectic operators can be defined. The one to update the particle position is referred to {\it Drift} and the other one to update momentum is {\it Kick}. The specific expressions read
\begin{eqnarray}
D(t_1;t_2) &:& {\mathbf r}(t_2) = {\mathbf r}(t_1) + {\mathbf p} \int^{t_2}_{t_1} \frac{dt}{a^2}, \notag \\ K(t_1;t_2) &:& {\mathbf p}(t_2) = {\mathbf p}(t_1) - \nabla \phi \int^{t_2}_{t_1} \frac{dt}{a}.
\end{eqnarray}
The Draft and Kick must be alternately applied.  This can be implemented with two feasible updating schemes, Kick-Drift-Kick (KDK) or Drift-Kick-Drift (DKD). For the KDK scheme, the momentum is updated by a half time step (from  $t_1$ to $t_1+(t_2-t_1)/2$) (Kick), then the position is updated by an entire step from $t_1$ to $t_2$ (Draft), finally the momentum is updated again by the remaining half step (from $t_1+(t_2-t_1)/2$ to $t_2$) (Kick). The integration order of DKD is just reversed to the KDK.

\begin{figure}[htbp]
\centering
\includegraphics[width=0.85\textwidth]{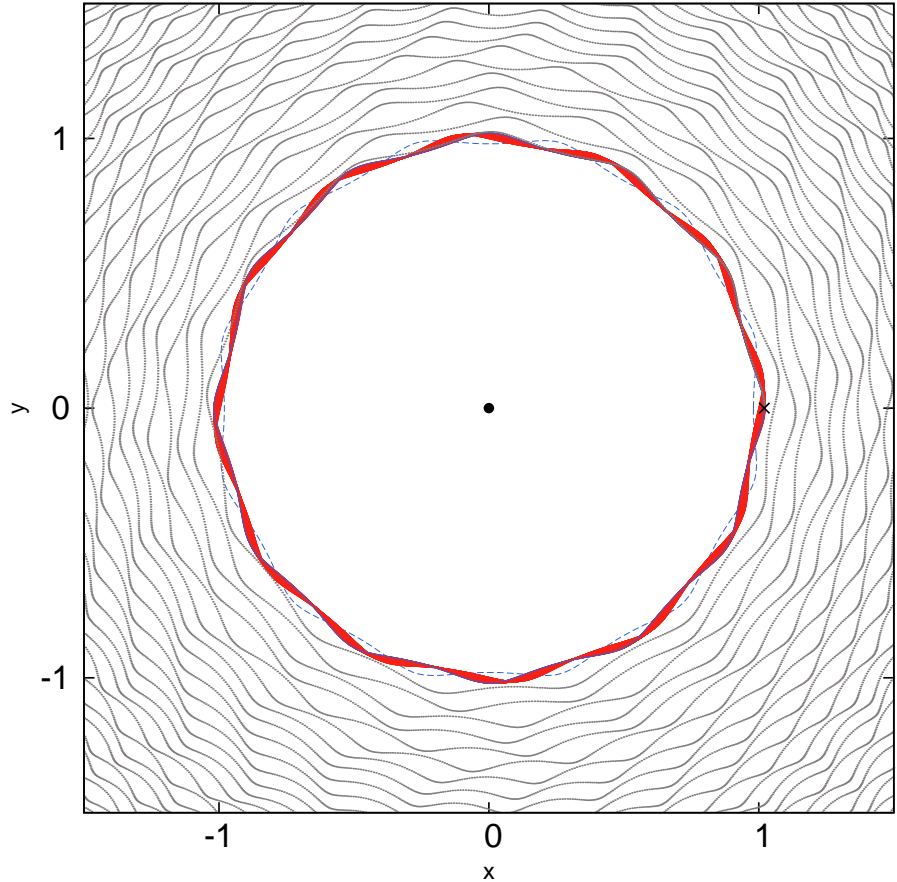} 
\caption{Adaptive KDK scheme. The central point is a massive point. The cross denotes initial position of one of close binary, the black solid curves denotes exact its closed orbit (nearly overlapping with the red curve) run with sufficiently small time step. The red curves denote KDK scheme with an insufficiently small time steps and the dotted  (gray) curve denotes DKD scheme with the same time step. The result suggests that KDK has better ability to conserve energy during long time evolution for a N-body system.}
\label{fig:kdk}
\end{figure}

In order to examine the robustness of those schemes under the decomposition of long and short range force calculation scheme, we setup a 3-body system, in which we set up two mass points orbiting around a central massive object. Fine-tuning the mass ratio, velocities and configurations so that the scale of the revolved orbit about the central mass point is larger than the split scale and the close binary is dominated by the short-range force. The exact orbits can be solved by an extremely high resolution integration with a direct 3-body gravity, which is a closed orbit denoted by the (black) solid curve in Fig.~\ref{fig:kdk}. 
Then we decrease the stepping rate. In KDK scheme (the red curve), the lower stepping rate is still stable but cause an extra procession. However the trajectory of DKD (the gray dotted curve) is unstable and spiraling outwards. It suggests that KDK is more robust for scale splitting as well. So \Code also employs the KDK scheme.

In cosmological $N$-body simulation, as matter become more and more clustered, a higher stepping rate (or shorter step) is needed to accurately follow the more rapid change of trajectory. \Code follow the criteria of Gadget-2,
\begin{equation}
t_{acc}=\frac{2\eta\epsilon}{\sqrt{|{\mathbf a}|}},  \notag
\end{equation}
to evaluate whether to refine the step length. If it does not be satisfied, the current step length is divided by 2, so that the step length of any particle in variant environment always has a 2 power of the top level. Such a half-and-half refinement is flexible to synchronize the adaptive steps among the particles in variant environments at different levels. 

\section{Implementation and Parallelization}

\subsection{Domain decomposition}

\label{sec:domain}

The essential task for parallelization of N-body problem on scalar architectures is how to distribute particles into individual processors. We follow the scheme proposed by \citet{2005MNRAS.364.1105S} with some improvements. The basic idea of the scheme is to use a self-similar Peano-Hilbert curve to map 3D space into a 1D curve which can be cut further into pieces that defines the individual domain. In practice, we firstly partition the simulation box into a  $2^{3n}$ grids and label each grid with a unique integer key, which is referred to as PH key, labeling the order of a point in the curve. Next, we sort the grids in order of the key and find the cutting positions of the filling curve, then we collect the particles in the same segment as a domain.  In our implementation, we follow approach of \citep{2012arXiv1211.4406I} to partition the 1D Curve according to the workload by using wall clock time of each segments as the weight to assess the workload in those segments. As shown by \citep{2012arXiv1211.4406I}, this can significantly improve load imbalance. Note, while the Peano-Hilbert curve is not the only option, but its decomposition is relatively spatial compact and has a low ratio of surface to volume, the latter has advantage in the communications of the domain boundaries.

\begin{figure}[htbp]
\centering
\includegraphics[width=0.75\textwidth]{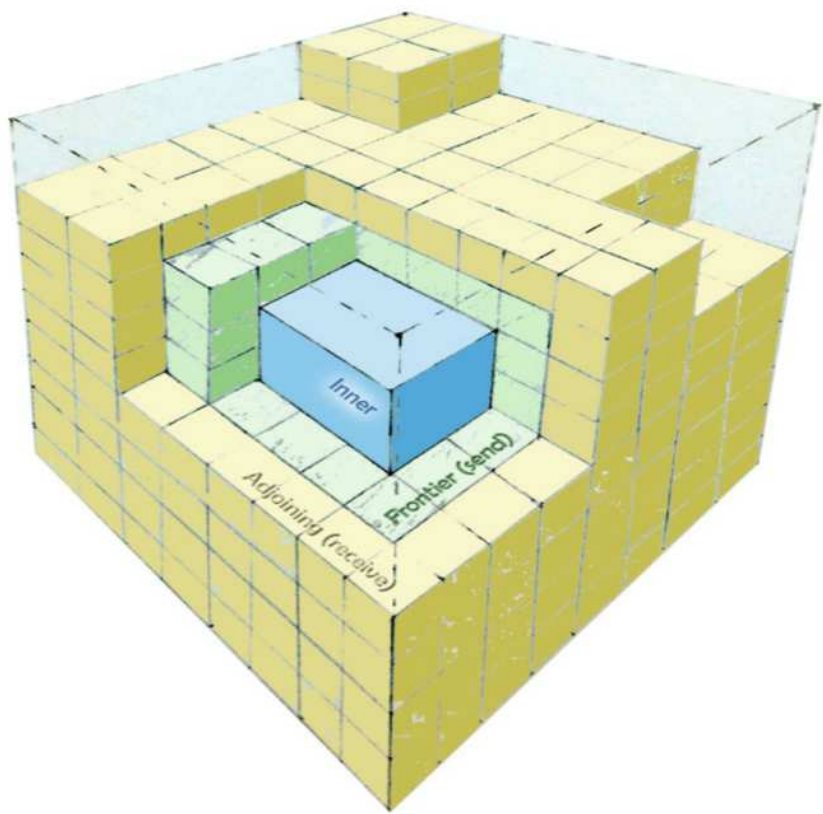}
\caption{A sketch of a domain in \Code. The domain in a computing node is a simple connected region determined by Peano-Hilbert filling curve method. A domain is surrounded by the ghost layer (yellow). The frontier part of the domain (green) needs to export data to its adjacent domains. The inner (blue) part is independent on other domains. The subcube denotes $ground$ nodes.}
\label{fig:domain}
\end{figure}

As an example, Fig.~\ref{fig:domain} illustrates the structure of a domain. Each grid is one of $2^{3n}$ described above, and a group of them make up of a domain. Since the short range force needs information from particles in adjacent domains, an entire domain includes a ghost layer containing data of adjacent domains. In Fig.~\ref{fig:domain}, the ghost layer( ``adjoining'' in the label) (yellow) stores information of the particles in its adjacent domains. Correspondingly, frontier layer (green) also needs to send its interior particle informations to its adjacent domains. Therefore we allocate extra buffer to store information for adjacent ghost layers, while the inner part (blue) of domain does not require informations from other domains. 

\subsection{Particle Mesh implementation}
\Code employs Particle-mesh method to compute the long-range force. It involves Fourier transformation which we use the public available FFTW package in \Code. However the data storage for FFTW does not match the strategy of our domain decomposition. In the FFTW, the mass density field must be assigned into the slabs along a specific direction.  In addition to the conventional {\bf MPI\_COMM\_WORLD} for the domain decomposition, we add an exclusive MPI communicator, {\bf PM\_COMM\_WORLD} for the PM.  The rank of {\bf MPI\_COMM\_WORLD} increases along the Peano-Hilbert key, while the rank of {\bf PM\_COMM\_WORLD} increases along the z direction. We construct the density mesh for the PM on the local domain, send the mesh information to the proper rank, then fill the information into the correct position.

\Code forks one thread for the PM and one thread for domain communications. The most time consuming part of the calculation--tree-walk, use all the remaining resource on each computing node. For instance, a CPU with 12 cores has to create a thread for domain decomposition, then create a thread for the PM and a thread for the Tree-building at the same time. Since PM calculation decouple with the Tree walking part, the PM and Tree traversal can be carried out at the same time, this improves scalability of the code because of the scalability of the FFT decreases with the number of employed meshes. 

\subsection{Task queues}
\begin{figure}[htbp]
\centering
\includegraphics[width=0.95\textwidth]{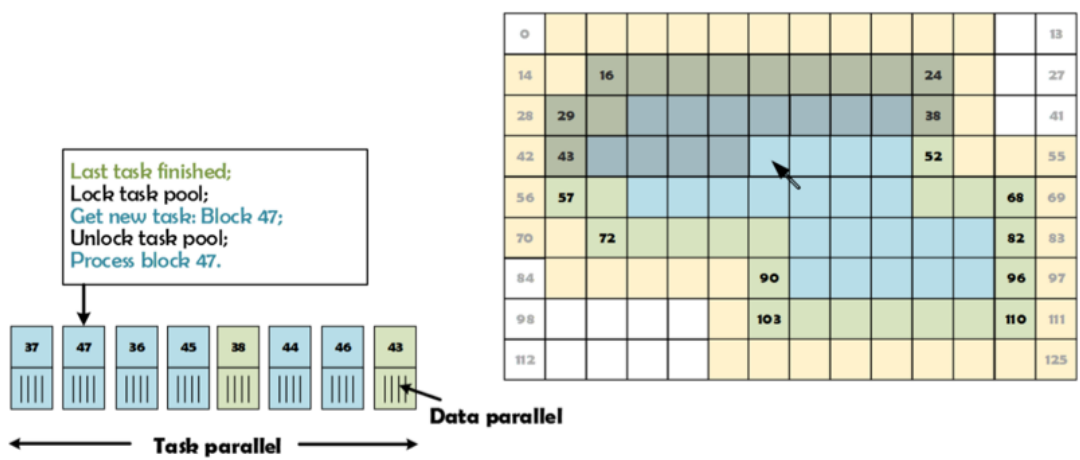}
\caption{The sketch of the task queue. The index label the order of ground grids (right panel) and it form a nature queue of tasks (left queue). Multiple threads roll the task queue in ascending order. In each task, \Code is optimized by the strategy of data parallel. 
}
\label{fig:queue}
\end{figure}

In our framework, the tree walking does not start from the root but from $ground$ grids. Each ground grid is a branch of local trees. Since the grid size is  exactly 6 times larger than the PM grid (see Fig.~\ref{fig:split}), each grid will complete the short-range force calculation by walking through the 26 surrounding adjoined local grids and itself. The right panel of Fig.~\ref{fig:queue} illustrates a domain consisting of the ground tree node. It is a nature queue pool to identify one grid as one task. It is an effective task-parallel strategy that multiple threads roll  the task queues, e.g. each thread fetches a grid data for the tree walking and the PP calculation. Such a task strategy can also better take advantage of the DTT method~\citep{2011arXiv1110.2921Y,koyama2013}.

\subsection{Heterogeneity}
\Code is designed for heterogeneous system. In this work, we focus on the Intel co-processor MIC heterogeneous architecture, whose advantage is high concurrency and wide SIMD (Single Instruction Multiple Data) instruction. Since the average bandwidth and memory access latency on MIC core is relatively lower than that of CPU. We offload particles onto MIC memory to accelerate the short-range force calculation. Because the particles are recorded on the $ground$ grain, the computing workload can be readily estimated. After assessing the particles numbers in grains, we can determine an optimized partition workload ratio between CPU and MIC in task queues. \Code also provide a dynamic mechanism to adjust the partition ratio according to the real elapsed time on MIC and CPU. The MIC run with offload mode and the data of the particles and tree nodes is transferred to MIC memory at beginning of each time step. This can optimizes communication between MIC and main memory and avoid repeated data exchange.

The forces for particles in a leaf are computed with direct summation method PP. To speed up this procedure, we optimize its performance with the idea of data parallelism. This allows us put more particles into leafs without additional time consuming.  This substantially reduces depth of the tree and so decreases the tree- walk. According to our experiment, the maximum particle number in a leaf can be set between 512 and 2048 on MIC and between 16 to 128 on CPU \citep{2013hpcn.confE...6H}.

\section{Test and Performance}
\label{sec:exam}
\begin{figure}[htbp]
\centering
\includegraphics[width=0.47\textwidth]{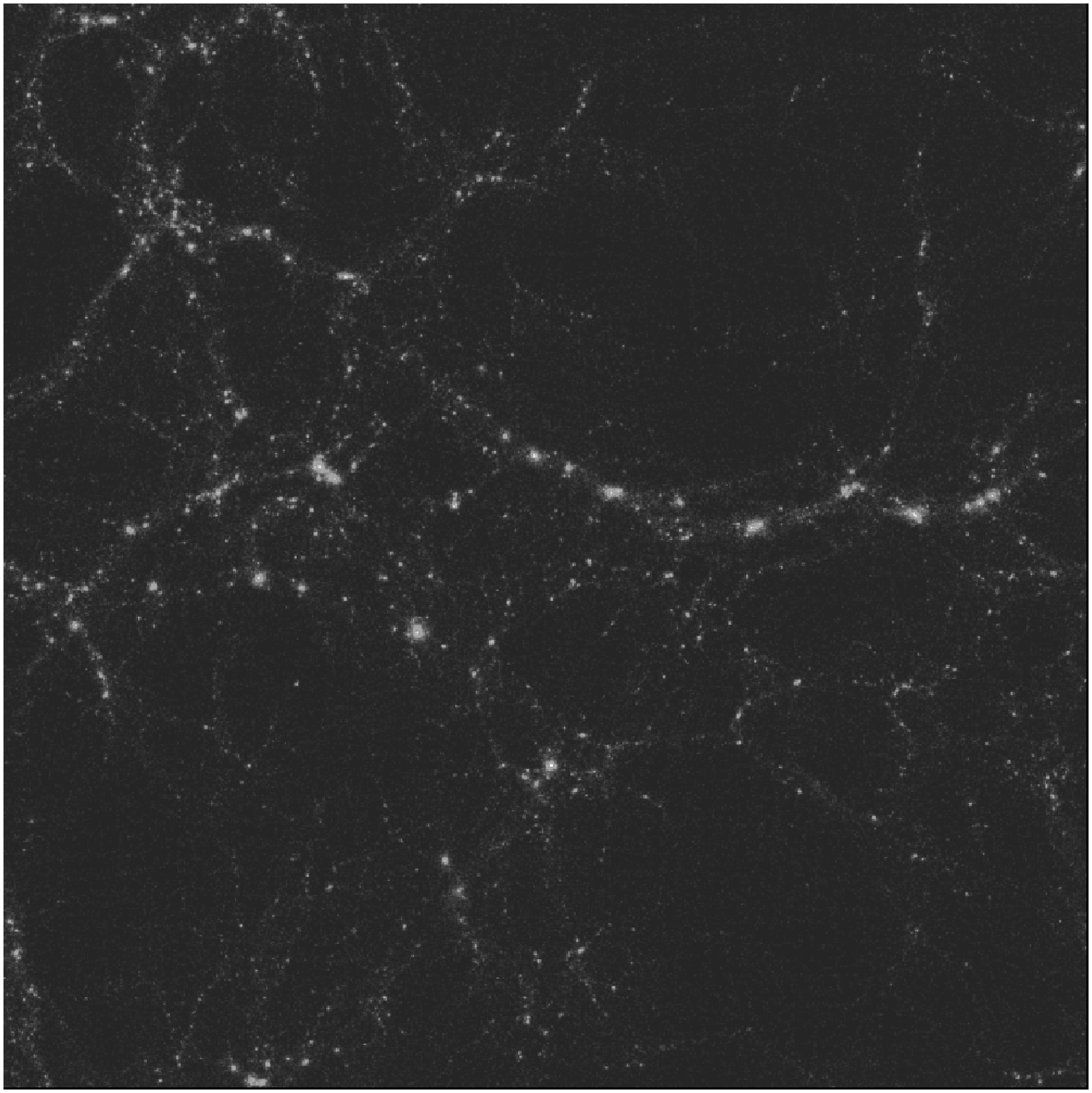}
\includegraphics[width=0.47\textwidth]{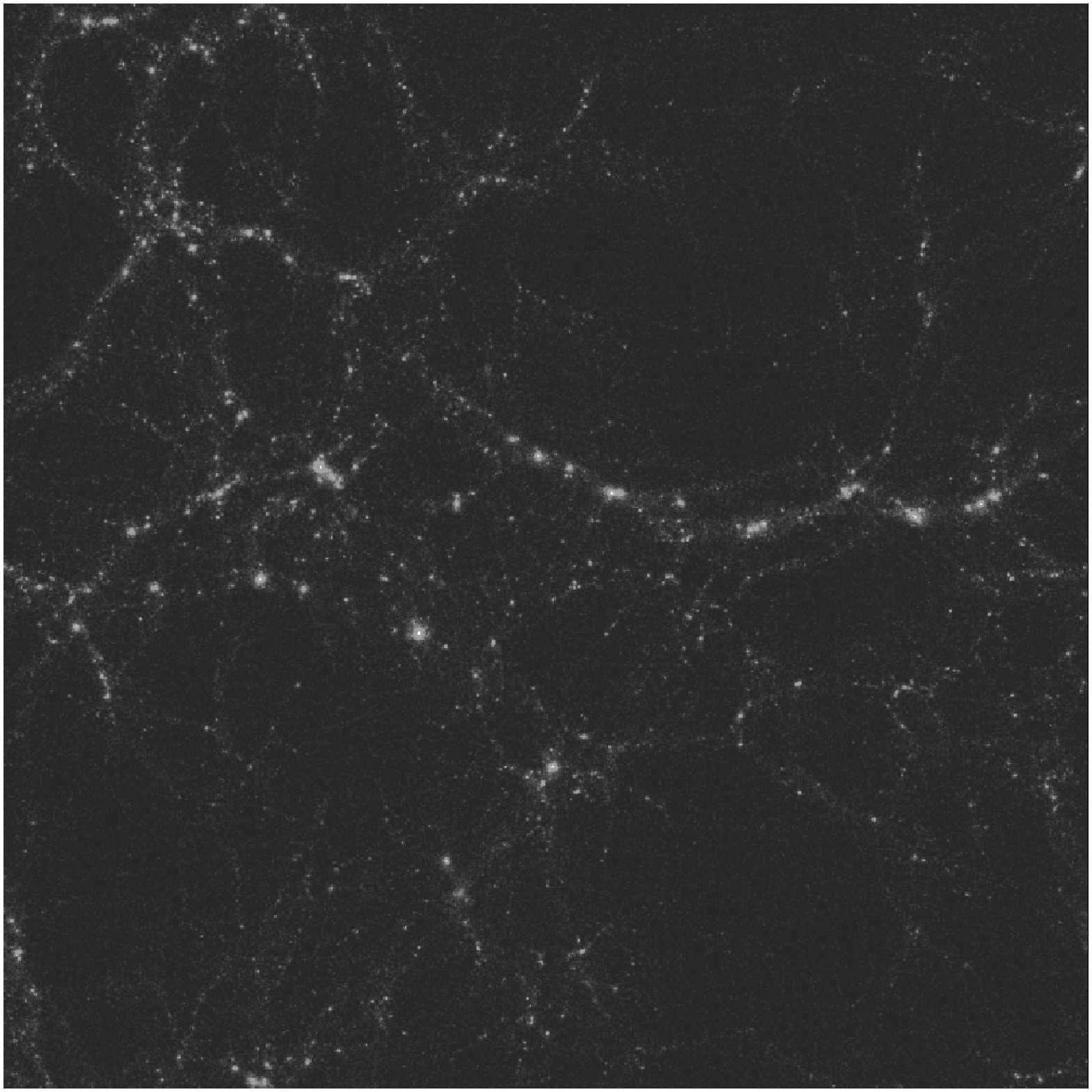}
\caption{Comparison of density map of our simulations run with different code. The left panel shows a slice of the simulation run with Gadget-2, the right panel shows result from \Code.}
\label{fig:snap}
\end{figure}

In order to test the accuracy of our code we carry out two cosmological N-body simulations starting from an identical initial condition but run with \Code and Gadget which has been commonly used in the community. The Simulations follow $128^3$ particles within a comoving box with a side length 100 $h^{-1}$Mpc. The initial condition of our simulations is generated by 2LPTic at redshift $z=49$, and the cosmological parameters are assumed to be $\Omega_M = 0.25, \Omega_{\Lambda} = 0.75, H_0 = 0.7, \sigma_8 = 0.8$. 
In Fig.~\ref{fig:snap}, we provide visualization of the matter density map of both simulations at $z=0$. Clearly two maps are indistinguishable. 

\begin{figure}[htbp]
\includegraphics[width=0.32\textwidth]{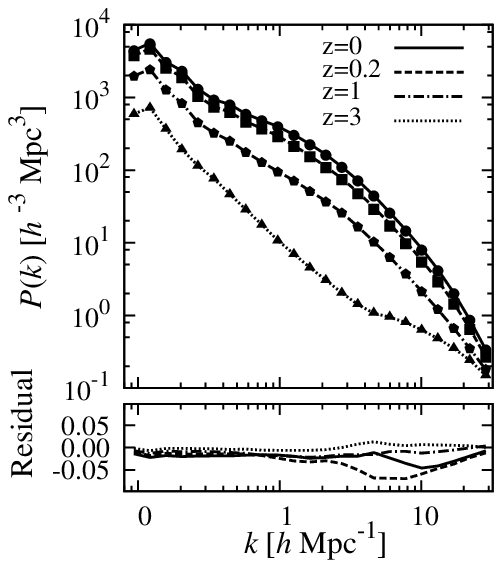} 
\includegraphics[width=0.32\textwidth]{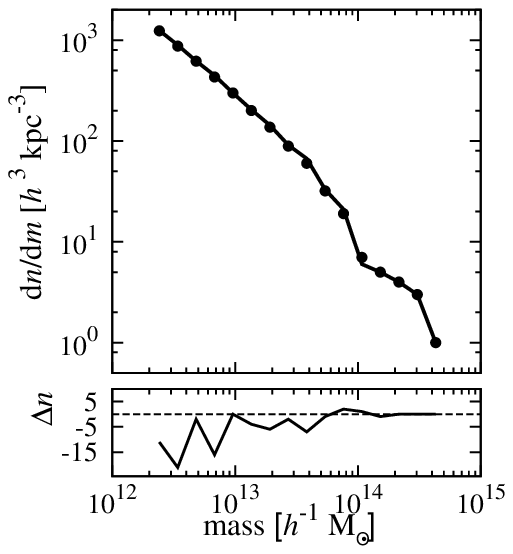} 
\includegraphics[width=0.32\textwidth]{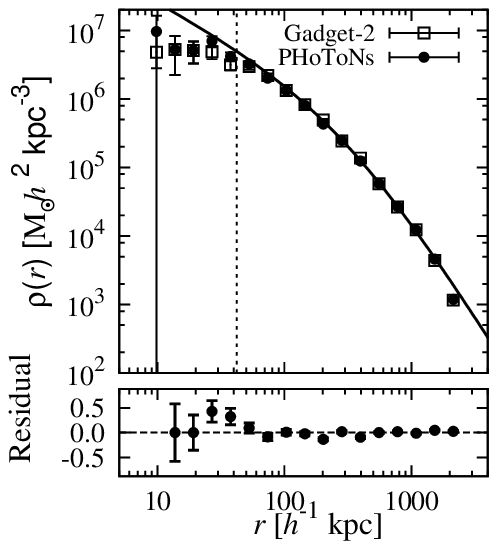}
\caption{Comparisons of various properties from simulations run with Gadget-2 and \Code, power spectrum (left panel), dark matter halo mass function (middle panel) and density profile of the largest halo in each simulation (right panel). In the left and the middle panels,  lines indicate results from Gadget and various symbols are from \Code. In the right panel, the open squares are from Gadget-2 and the filled squares are from \Code, as indicated in the label. $\Delta n$ is the difference of halo number in each mass bin. }
\label{fig:pow}
\end{figure}

In figure~\ref{fig:pow} we provide quantitatively comparison of both simulations. In left panel of the figure, we compare matter power spectrum of both simulations at different epoch, z=0, 0.2, 1 and 4. Solid lines are results from the run with \Code and symbols are from the run with Gadget. As can be seen clearly, matter power spectrum in both simulations are identical. The middle panel of the figure present comparison of halo mass function in both simulations. Again results from two simulation are indistinguishable. In the right panel of the figure, we plot density profile of the most massive halo in both simulations. The vertical dashed lines shows the softening length our simulations, and the solid curve is the best NFW fit to the profiles. Clearly, density profiles of both simulations agree well each other down to the softening length . This suggests that the numerical accuracy of the \Code is comparable to the popular code--Gadget~\citep{2008CS&D....1a5003H,2014ApJS..210...14K}.

Our benchmark runs on $Era$ computer \footnote{http://www.sccas.cn/yhfw/yjzy/xyd/} of 300 Tflop/s (Double Precision) cluster. The computer contains 256 CPU nodes, 40 MIC nodes and 30 GPU nodes. Each node comprises two twelve 2.8GHz Ivy Bridge Xeon E5-2680v2 cores and 64/128 GB DDR3 system memory. These nodes are inter-connected with the EDR Infiniband which provides 56  Gbps peer to peer bandwidth. Each of the 40 MIC codes has two Xeon Phi 5110p cards with 60 cores (4 threads per core) running at 1.05 GHz and 8GB GDDR5 memory. The peak SP (single precision) performance is 896 Gflops for CPU node and 4.926 Tflop/s for MIC node.

\begin{table}
\centering
\caption{\label{tab:scalability}The scalability of \Code on $Era$ super computer. $N_{side}$ denotes the number of PM meshes used in tests. CPU and C+M indicate wall-clock time run with pure CPUs and CPU+MIC, respectively.}
\begin{tabular}{crrrrrrr}
\#Node &  2 & 4 & 8 & 16 & 24 & 32 & 40 \\
\hline
$N_{side}=192$\\
CPU(s) & 721.0 & 364.0 & 188.4 & 97.7 & 66.7 & 51.6 & 42.0 \\
C+M(s) & 240.9 & 123.1 &  66.4 & 34.2 & 24.8 & 19.4 & 17.5 \\
\hline
$N_{side}=384$\\
CPU(s) & 270.8 & 137.5 &  71.9 & 38.7 & 28.3 & 21.4 & 20.3 \\
C+M(s) &  54.9 &  30.9 &  17.3 & 11.0 & 10.2 &  8.2 &  9.5 \\
\end{tabular}
\end{table}

Table~\ref{tab:scalability} shows that \Code has good scalability both on pure CPU and CPU+MIC platforms. Upon the limit of our test machine, 40 nodes on $Era$, the wall-clocking time roughly decreases linearly. CPU+MIC architecture generally speeds up the CPU platform by roughly 3 times. Assuming 22 flops in one interaction~\citep{2006NewA...12..169N}, the efficiency goes up to 68.6\% of peak performance on MIC and 74.4\% on two Sandy Bridge CPU.

\section{Summary}
\label{sec:discussion}
We describe a new cosmological $N$-body simulation code \Code. The code is designed to run on both pure CPU based and Heterogeneous platforms. In particularly, our implementation on heterogeneous platform is dedicated to perform massive simulation on the CPU+MIC architecture, for instance, Tianhe-2 supercomputer.

\Code adopts a hybrid scheme to compute gravity solver, including the conventional PM to calculate the long-range force and the Tree method  for short-range force, and the direct summation PP to compute force between very close particles. A merit of the \Code is to better take advantage of multi-threads feature of new generation of supercomputer by using Dual Tree Traversal and a flexible task queue to make use of multiple threads and improve the load imbalance. In addition, our PM computation is independent of our tasks and thus can be hidden during the tree-walk, which improves the scalability of code. The performance is effectively improved by optimizing the PP kernel with data parallelism on heterogeneous architecture of MIC. The future development of the \Code will include hydrodynamics, which will be presented in the other paper.

\normalem
\begin{acknowledgements}
We acknowledge supports from the National Key Program for Science and Technology Research and Development (2017YFB0203300) and  NSFC grants (11403035, 11425312 and 11573030). LG acknowledges support from Royal Society Newton advanced Fellowships.
\end{acknowledgements}

\bibliographystyle{raa}
\bibliography{ms}

\end{document}